\documentclass[twocolumn,tighten,twocolappendix]{aastex631}
\usepackage{amsmath}
\usepackage{hyperref}

\def\mw{{Milky\,Way}~}
\def\mwn{{Milky\,Way}}

\def\hi{{{\rm H}\,{\sc i}~}}
\def\hin{{{\rm H}\,{\sc i}}}
\def\ovii{{{\rm O}\,{\sc vii}~}}
\def\oviin{{{\rm O}\,{\sc vii}}}
\def\oviii{{{\rm O}\,{\sc viii}~}}
\def\oviiin{{{\rm O}\,{\sc viii}}}
\def\oviia{{{\rm O}\,{\sc vii}~{\rm K$\alpha
$}~}}
\def\oviib{{{\rm O}\,{\sc vii}~{\rm K$\beta
$}~}}
\def\oviian{{{\rm O}\,{\sc vii}~{\rm K$\alpha
$}}}
\def\oviibn{{{\rm O}\,{\sc vii}~{\rm K$\beta
$}}}

\def\chandra{{\it Chandra}~}

\def\xmm{{\it XMM-Newton}~}
\def\xmmn{{\it XMM-Newton}}

\shorttitle{Non-thermal broadening}
\shortauthors{Das et al.}

\begin{document}

\title{Detecting the effect of non-thermal sources on the warm-hot Galactic halo}

\correspondingauthor{Sanskriti Das}
\email{snskriti@stanford.edu}

\author[0000-0002-9069-7061]{Sanskriti Das}
\affil{Kavli Institute for Particle Astrophysics and Cosmology, Stanford University, 452 Lomita Mall, Stanford, CA\,94305, USA}

\begin{abstract}
We report the first detection of non-thermal broadening of \ovii lines in the warm-hot $\approx 10^6$\,K circumgalactic medium (CGM) of the Milky\,Way. We use $z$=0 absorption of \oviian, \oviibn, and \oviii K$\alpha$ lines in archival grating data of $b>$15$^\circ$ quasar sightlines from \chandra and \xmmn. Non-thermal line broadening is evident in 2/3rd of the sightlines considered, and on average is constrained at 4.6$\sigma$ significance. Non-thermal line broadening dominates over thermal broadening. We extensively test 
whether the appearance of non-thermal line broadening could instead be because of multiple thermally broadened velocity components and robustly rule it out. Non-thermal line broadening is more evident toward sightlines at lower galactic latitude indicating the Galactic disk origin of the nonthermal sources. There is weak/no correlation between non-thermal line broadening and the angular separation of sightlines from the Galactic center, indicating that the nuclear region might not be a major source of non-thermal factors.
\end{abstract}


\section{Introduction} \label{sec:intro} 
The circumgalactic medium (CGM) is the multiphase gaseous region surrounding the disk of a galaxy and filling up its dark matter halo \citep{Truong2023}. The most massive and volume-filling component of the CGM is expected to be virialized, warm-hot, and diffuse \citep[$\approx 10^6$\,K;][]{Spitzer1956,Fielding2020}. This warm-hot $\gtrapprox 10^6$\,K gas is observed using the emission and absorption lines of He-like and H-like metal ions, e.g., \ovii and \oviiin, and free-free continuum emission in X-ray. However, due to the intrinsic difficulties in detecting the warm-hot gas, the warm-hot CGM has been best characterized for the \mw  \citep[see the review by][]{Mathur2022}. 

\begin{figure*}
    \centering
    \includegraphics[width=0.325\textwidth]{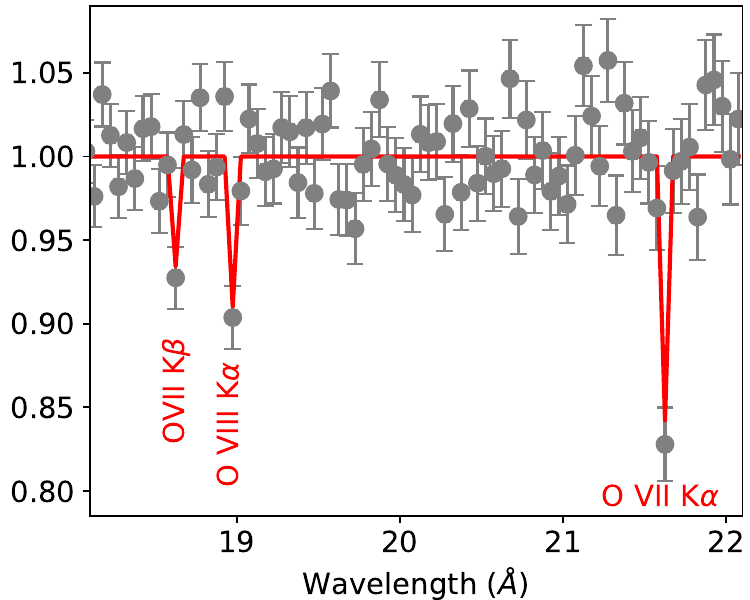}
    \includegraphics[width=0.325\textwidth]{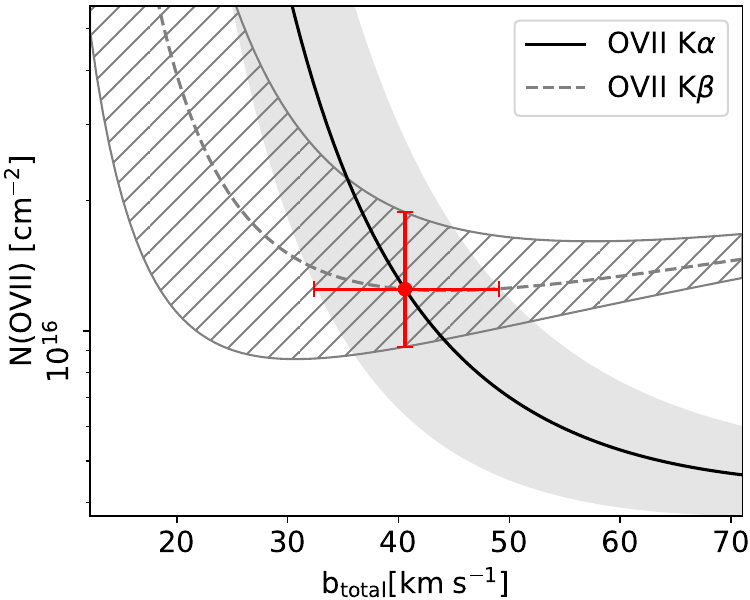}
    \includegraphics[width=0.325\textwidth]{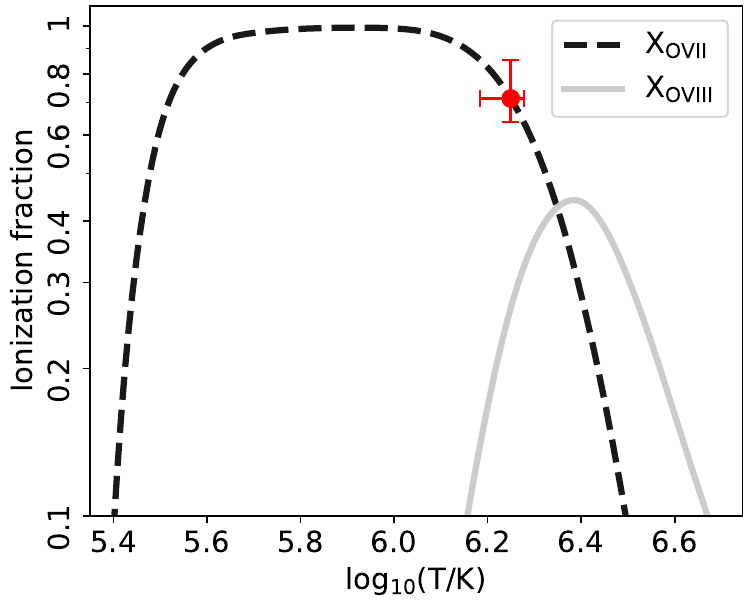}
    \caption{Left: Normalized unfolded spectrum toward one of the sightlines. The $z=0$ absorption lines of \oviian, \oviibn, and \oviii K$\alpha$ are shown. Middle: Contours of N(\oviin) and total broadening of \oviin, $\rm b_{total}$ for the measured equivalent widths of K$\alpha$ and K$\beta$ lines toward the same sightline. The saturation-corrected N(\oviin) and corresponding $\rm b_{total}$ are shown with the red point. Right: Ionization fraction of \ovii and \oviiin, $\rm X_{OVII}$ and $\rm X_{OVIII}$, as a function of temperature in collisional ionization equilibrium. The temperature and $\rm X_{OVII}$ toward the same sightline {obtained from the column density ratio of \oviii and \ovii (see equation\,\ref{eq:getT})} is shown with the red point.}
    \label{fig:get_b}
\end{figure*}

CGM is expected to be influenced by non-thermal sources, e.g., turbulent motions, cosmic rays, and magnetic fields. Under the assumption of hydrostatic equilibrium, non-thermal sources enhance cooling in the hot gas resulting in the generation of cooler and denser phases to balance the total (thermal and non-thermal) pressure against gravity \citep{Bennett2020,Ji2020,Schmidt2021,Voort2021}. Thus the presence of non-thermal sources in the warm-hot CGM is a natural indicator of coexistent phase(s) at lower temperatures. However, directly constraining the strength of non-thermal sources in the CGM is extremely challenging. For example, nonthermal emission from cosmic-ray electrons in 144\,MHz and magnetic fields using Faraday rotation have been detected in the inner CGM of nearby galaxies \citep{Heesen2019,Heesen2023}, but whether it affects the warm-hot phase is unclear. One observable effect of non-thermal sources is that the velocity dispersion is enhanced and the line broadening of tracer element transitions (e.g., \ovii in the warm-hot CGM) is larger than their thermal broadening.   

In previous studies of the warm-hot CGM in X-rays, the focus has been to detect the warm-hot $\gtrapprox 10^6$\,K CGM and identify it as the CGM (i.e., ruling out the sources of confusion), calculate the temperature, column density in absorption \citep[e.g.,][]{Nicastro2002,Williams2005,Fang2015,Das2019a,Das2021b}, emission measure in emission \citep[e.g.,][]{Henley2010,Das2019c,Gupta2023}, and estimate the baryonic mass \citep{Gupta2012,Nicastro2016b}. In this letter, we take a step forward and attempt to constrain the non-thermal line broadening in the warm-hot CGM.

The letter is structured as follows. We present the analysis in \S\ref{sec:method}, show the results, and discuss the physical implications in \S\ref{sec:resdis}. We summarize our results and discuss future directions in \S\ref{sec:summary}.

\section{Method}\label{sec:method}
Our focus is on the warm-hot CGM of the Milky Way. {Because oxygen is the most abundant metal in solar-like chemical composition, and \ovii and \oviii are the dominant oxygen ions at the temperature of warm-hot phase}, we consider the wavelength regions around $z=0$ transitions of \oviian, \oviibn, and \oviii K$\alpha$ at 21.602\AA, 18.627\AA, and 18.978\AA. To minimize the contamination by the interstellar medium (ISM), we consider extragalactic quasar sightlines at high galactic latitudes ($b$$>$15$^\circ$). We exclude the targets whose blueshifted intrinsic warm absorbers could overlap with $z=0$ lines of our interest. From the \chandra archive, we extract all grating data of consideration: 25 HRC/LETG, 14 ACIS/LETG, and 60 ACIS/HETG-MEG sightlines\footnote{We exclude ACIS data after cycle 14 because of its degradation in soft X-ray}, and reduce them with CIAO following standard procedure \citep[see \S2.1 of][]{Das2021b}. We also consider the quasar sightlines in \xmm RGS data published in \cite{Nicastro2016b}. 

To make sure that the \oviia line is detectable, we exclude the sightlines with signal-to-noise ratio per resolution element, SNRE$<$10 around the \oviia line. In XSPEC, we fit the continuum with a power law absorbed by the Galactic \hi (\texttt{tbabs*powerlaw}) and model the oxygen absorption lines with unresolved Gaussian (\texttt{agauss}). We freeze N(\hin) toward each sightline according to \cite{Bekhti2016}. We allow the wavelength of the Gaussian lines to vary within the resolution element (50\,m\AA~for LETG, 25\,m\AA~for MEG, 70\,m\AA~for RGS) around the expected $z=0$ values.  

To calculate the thermal broadening, we need to constrain the temperature from \oviii and \oviin. Therefore we exclude sightlines where \oviii K$\alpha$ lines are not detected, i.e., where the thermal broadening cannot be estimated.  {Because the oxygen lines are unresolved, we cannot \textit{directly} measure the total (thermal+non-thermal) line broadening.} To \textit{indirectly} calculate the total broadening, both \oviia and \oviib lines are required. Therefore we exclude sightlines where \oviib lines are not detected, i.e., where the total broadening cannot be estimated. 

If \ovii lines are unsaturated, the \ovii column density, N, is independent of its total line broadening \citep{Draine2011}. Thus the equivalent width (EW) ratio of \oviib and \oviia depends only on the {rest-frame} transition wavelength, $\lambda$, and the oscillator strength\footnote{{A dimensionless quantity that is a measure of the \href{https://www.nist.gov/pml/atomic-spectroscopy-compendium-basic-ideas-notation-data-and-formulas/atomic-spectroscopy}{probability of the transition}}}, $f$: 
\begin{equation}
\begin{split}
    {\rm N} = \frac{\pi e^2}{m_e c^2} \frac{\rm EW}{f\lambda^2}\\
    \frac{\rm EW_{OVII\,K\beta}}{\rm EW_{OVII\,K\alpha}} = \frac{f_{\rm OVII\,K\beta}\lambda_{\rm OVII\,K\beta}^2}{f_{\rm OVII\,K\alpha}\lambda_{\rm OVII\,K\alpha}^2}
\end{split}
\end{equation}
In this scenario, we cannot {estimate the total line broadening and therefore exclude} those sightlines for our analysis. We end up with 12 sightlines\footnote{{Mrk\,421 and PKS\,2155-304 (ACIS-S/LETG); Mrk\,509 and NGC\,5548 (HRC-S/LETG); 3C\,282, Mrk\,290, and NGC\,3783 (ACIS-S/HETG-MEG); 1ES\,1553+113, 3C\,273, 3C\,390.3, HE\,1029-1401, and PKS\,2005-489 (RGS)}}, 7 from \chandra and 5 from \xmmn, where \oviian, \oviibn, and \oviii K$\alpha$ lines are detected (Fig.\,\ref{fig:get_b}, left panel), and the equivalent width ratio of \oviib and \oviia is larger than expected from an unsaturated line, indicating saturation of the \oviia line. 

\begin{figure*}
    \centering
    \includegraphics[width=0.95\textwidth]{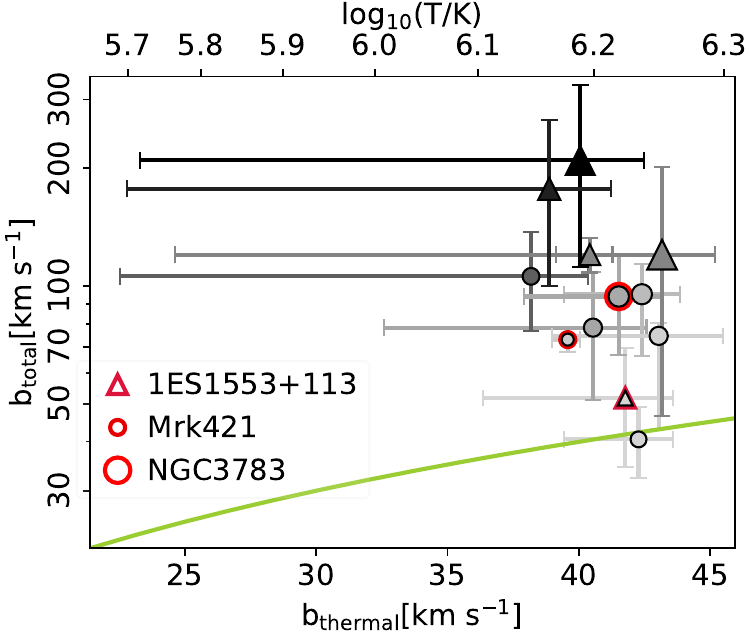}
    \caption{Comparison between thermal and total broadening of \ovii lines. The temperature corresponding to the thermal broadening is shown in the upper x-axis. The shade and size of the symbols are proportional to N(\oviin) and N(\oviiin), respectively, with darker symbols having larger N(\oviin) and larger symbols having larger  N(\oviiin). Measurements from \chandra and \xmm are shown with circles and triangles, respectively. The green curve denotes the total broadening expected in the absence of non-thermal broadening. In 8 out of 12 sightlines non-thermal broadening is evident. Three sightlines where the presence of a super-virial {$\approx 10^7$\,K hot} phase coexisting with the virial {$\approx 10^6$\,K warm-hot} phase is known, are highlighted in red. {The true non-thermal broadening of \ovii lines toward these sightlines might be larger than what we measure here in a single phase scenario (see \S\ref{sec:multiT} for details)}.}
    \label{fig:vturb}
\end{figure*}

We determine the column density \citep[similar to previous works, e.g.,][]{Williams2005,Gupta2012} and the total line broadening of \ovii as follows. For a saturated but undamped line transition, the column density, N, depends on the equivalent width, EW, of that transition and the total broadening $\rm b_{total}$ \citep{Draine2011}:
\begin{equation}
{\rm N} =  \frac{\rm ln 2}{\sqrt{\pi}}\frac{m_e c}{e^2}\Big(\frac{\rm b_{total}}{f\lambda}\Big)exp\Big[\Big(\frac{c{\rm EW}}{2{\rm b_{total}}\lambda}\Big)^2\Big]
\end{equation}
We construct the contours of N(\oviin) vs $\rm b_{total}$(\oviin) for the measured EW values of \oviia and \oviibn. We estimate the saturation-corrected N(\oviin) and $\rm b_{total}$(\oviin) from where the contours intersect (Fig.\,\ref{fig:get_b}, middle panel). \oviii K$\alpha$ line is optically thin and unsaturated, so N(\oviiin) is linearly proportional to its EW.

The column density of the $i$-th ionization state of a metal M, $\rm M_i$, depends on the column density of hydrogen in that ion-containing phase, N(H), the abundance of that metal to hydrogen, $\rm A_{M/H}$, and the temperature-dependent ionization fraction of that ion, $\rm X_{M_i}$ (equation\,\ref{eq:X2N}). Thus, two different ions of the same metal can be used to estimate the temperature of the phase containing those metal ions. Applying this to oxygen, and assuming that all of the \oviii and \ovii are coming from a single phase, we can estimate the temperature, T, of that phase (equation\,\ref{eq:getT}; Fig.\,\ref{fig:get_b}, right panel) from the ratio of N(\oviiin) and N(\oviin).
\begin{subequations}
    \begin{equation}\label{eq:X2N}
    \begin{split}
    \rm N(OVII) = N(H)A_{O/H} X_{OVII}(T)\\
    \rm N(OVIII) = N(H)A_{O/H} X_{OVIII}(T)
\end{split}
\end{equation}
\begin{equation}\label{eq:getT}
    \rm\frac{N(OVIII)}{N(OVII)} = \frac{X_{OVIII}(T)}{X_{OVII}(T)} \rightarrow \; T
\end{equation}
\end{subequations}

From the temperature, T, we obtain the thermal broadening of oxygen lines {using equation\,\ref{eq:btherm}}: 
\begin{equation}\label{eq:btherm}
    {\rm b_{thermal}} = \sqrt{\frac{2k_B {\rm T}}{m_O}}
\end{equation}

\section{Results and discussion}\label{sec:resdis}

\begin{figure*}
    \centering
    \includegraphics[trim=5 3 3 10, clip, width=0.7\textwidth]{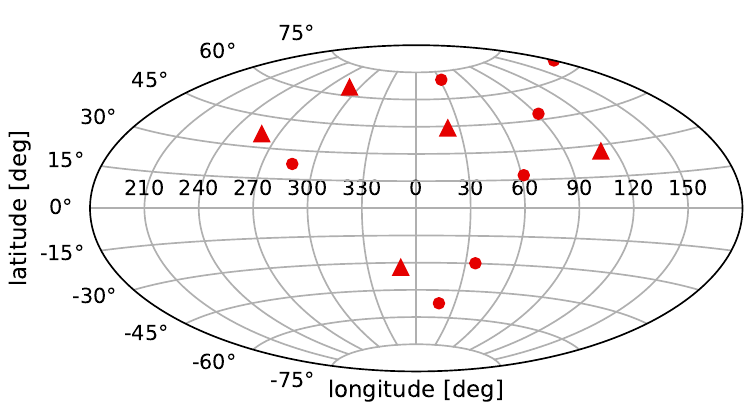}
    \gridline{
          \fig{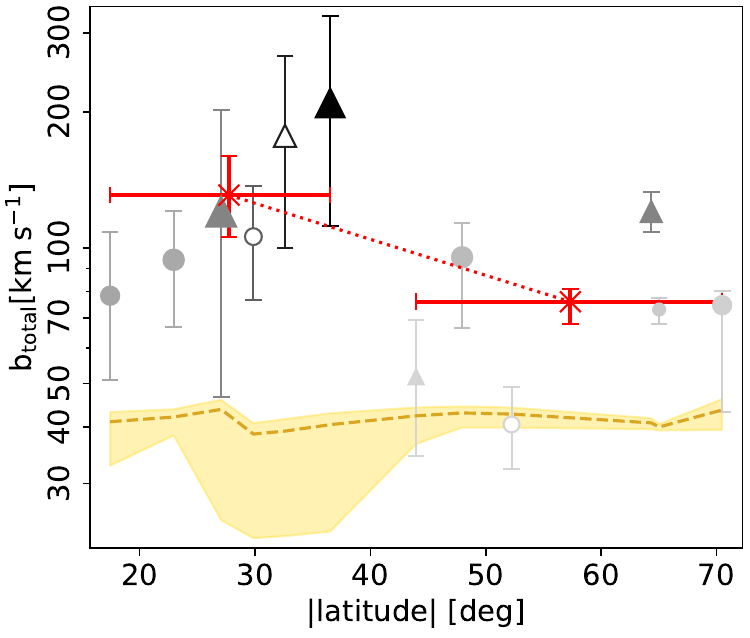}{0.325\textwidth}{\vspace{-0.2 in}(a)}
          \fig{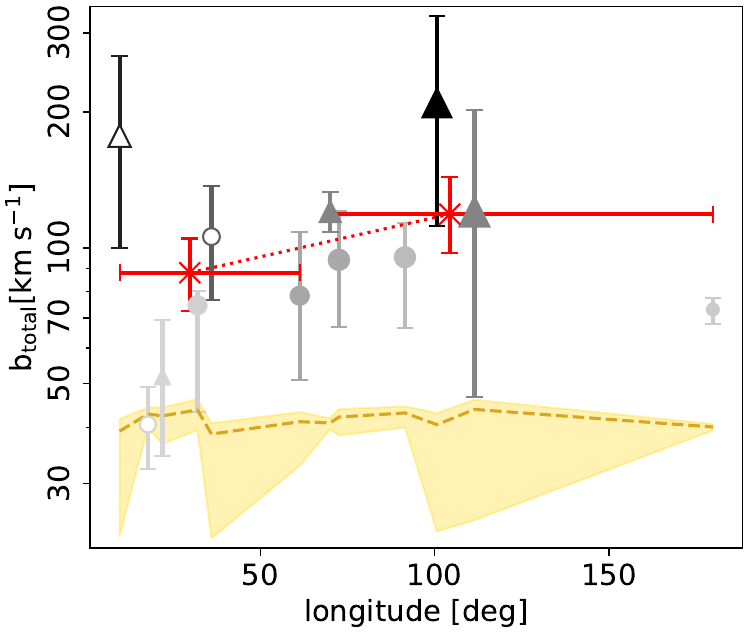}{0.325\textwidth}{\vspace{-0.2 in}(b)}
          \fig{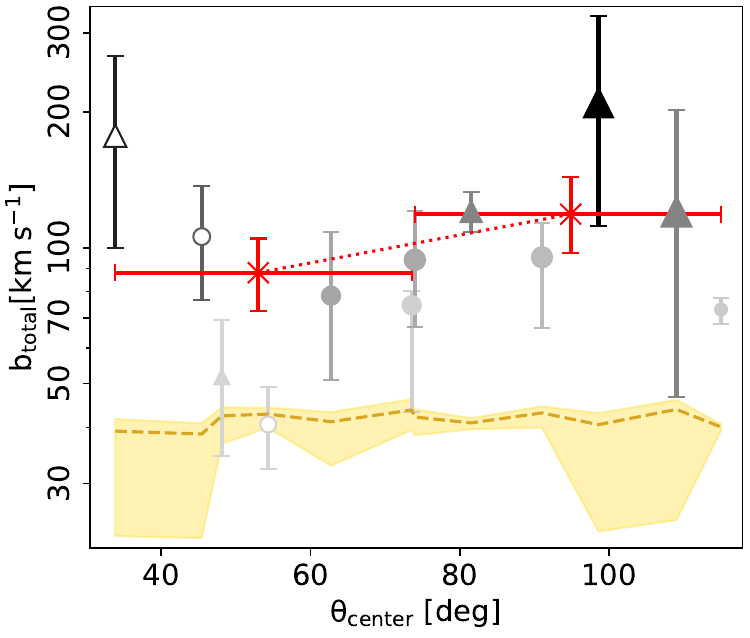}{0.325\textwidth}{\vspace{-0.2 in}(c)}}
    \vspace*{-.25 in}
    \caption{Top: Sky distribution of the sightlines, with \chandra and \xmm data shown with circles and triangles. Bottom: \ovii line broadening at different galactic latitudes (left), longitudes (middle), and angle from the Galactic center, $\theta_{\rm center}$ (right). The binned line broadening is shown with the red points. The grayscale symbols' size, shape, and shades are the same as Fig.\,\ref{fig:vturb}. Sightlines in the southern hemisphere ($b<$0) are shown with unfilled symbols. The thermal broadening is shown with the dashed brown curves and the yellow-shaded region.  Non-thermal line broadening is evident toward sightlines at lower galactic latitudes. The correlation between non-thermal line broadening and galactic longitude or $\theta_{\rm center}$ is weak/negligible.}
    \label{fig:bvssky}
\end{figure*}

We show the total broadening vs thermal broadening of \ovii toward all 12 sightlines in Fig.\,\ref{fig:vturb}. The large uncertainty in the temperature and hence the thermal broadening on the lower end is due to the flattening of the \ovii ionization fraction in the temperature range of 10$^{5.6-6.2}$\,K. The thermal broadening of all sightlines is similar, but the total broadening spans an order-of-magnitude range. In 8 out of 12 sightlines, the total broadening is larger than the thermal broadening including 1$\sigma$ error, indicating the evidence of non-thermal line broadening.

On average, the thermal and total line broadening are $41^{+1}_{-3}$\,km\,s$^{-1}$ and $103^{+15}_{-13}$\,km\,s$^{-1}$, respectively. It implies a 4.6$\sigma$ detection of mean non-thermal line broadening of $62^{+15}_{-13}$\,km\,s$^{-1}$. The non-thermal broadening is $1.5^{+0.5}_{-0.3}$ times stronger than the thermal broadening. 

\subsection{Sky distribution}
The sightlines we consider are widely spread across the sky (Fig.\,\ref{fig:bvssky}, top panel). None of our sightlines pass through the CGM of M\,31, Magellanic Clouds or Magellanic Stream. Therefore, the detection of non-thermal broadening is unlikely to be connected to a special structure in the halo; it could be a general characteristic of the CGM. 

We study the variation of \ovii line broadening as a function of galactic latitude, longitude, and angle from the Galactic center, $\theta_{\rm center}$ = cos$^{-1}$[cos($l$)cos($b$)] in Fig.\,\ref{fig:bvssky} (bottom panels). To better understand the trend, if any, we average the total broadening into 2 bins of the respective galactic coordinates (red points). The thermal broadening is $\approx$constant across all sky positions. Thus, any trend of total broadening would be equivalent to that of non-thermal broadening. 

Sightlines at smaller galactic longitudes and/or $\theta_{\rm center}$ (i.e., inner sightlines) probe the inner CGM that could potentially be affected by the nuclear activity of the Galaxy. Sightlines at larger galactic longitudes and/or $\theta_{\rm center}$ (i.e., outer sightlines) are less/unlikely affected by nuclear activities. At smaller galactic longitudes and $\theta_{\rm center}$, non-thermal line broadening is evident in 3 out of 6 sightlines, which increase to 5 out of 6 sightlines at large galactic longitudes and $\theta_{\rm center}$. The binned total broadening increases by 35\% from inner to outer sightlines but is constant within 1$\sigma$ error. Thus, we do not see any obvious non-thermal effect of (past) nuclear activities on the warm-hot CGM. 

Sightlines at lower galactic latitudes are more strongly affected by activities in the Galactic disk. At lower galactic latitudes, non-thermal line broadening is evident in 5 out of 6 sightlines, which decrease to 3 out of 6 sightlines at higher galactic latitudes. The binned total broadening increases by $74^{+64}_{-42}$\% from higher to lower latitudes. It is consistent with the fact the sources of non-thermal broadening likely originate in the disk.

\subsection{Decomposing broadening from velocity}
In this section, we discuss if our detection of non-thermal broadening is real, or whether it is a manifestation of multiple thermally broadened components at different velocities. 

\begin{figure*}
\renewcommand{\thefigure}{4a}
    \centering
    \includegraphics[width=0.95\textwidth]{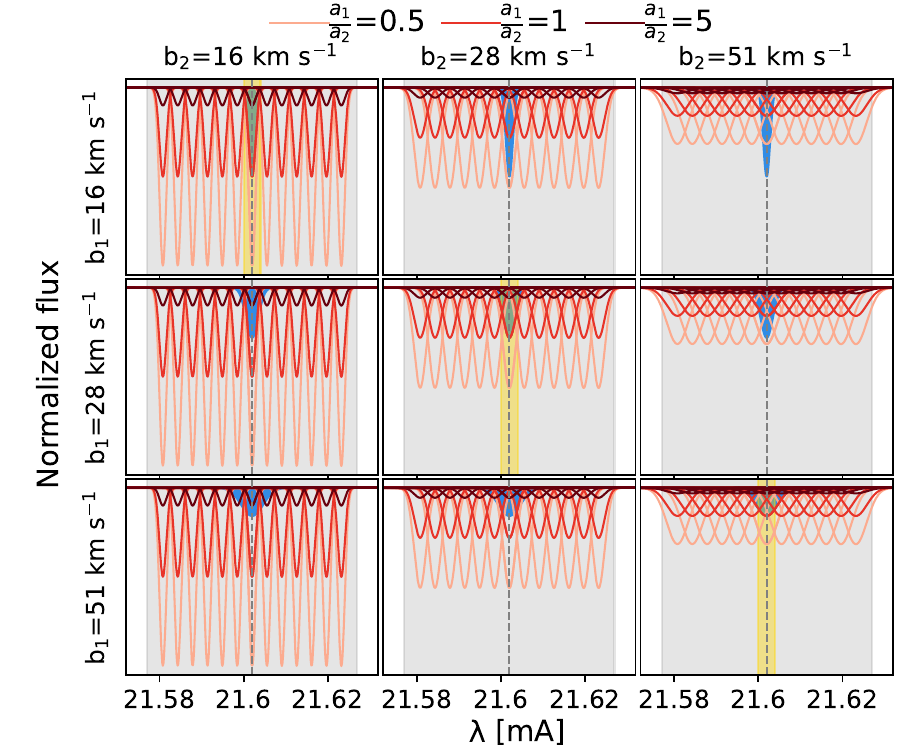}
    \caption{Illustration of the 2-Gaussian input models for \oviia in simulated spectra. The yellow-shaded regions denote the subset of models where 2-Gaussian effectively converges to 1-Gaussian. The broadening of input lines in each panel is labeled on the y-axis and upper x-axis. The first Gaussian is shown in blue. Different shades of red denote 3 different amplitudes of the second Gaussian (see the legend). The second Gaussian is shown for 13 line-of-sight (LOS) velocities. For each combination of line broadening, line amplitude ratio, and LOS velocity, the input is the first Gaussian and one of the 3$\times$13 red-shaded Gaussians. The vertical gray dashed line and the shaded region correspond to the $z$=0 wavelength of \oviia and the resolution element (of LETG), respectively. Models for \oviib would look similar except for the amplitude of all lines being weaker.}
    \label{fig:disentangle_sim}
\end{figure*}

\begin{figure*}
\renewcommand{\thefigure}{4b}
    \centering
    \includegraphics[width=0.95\textwidth]{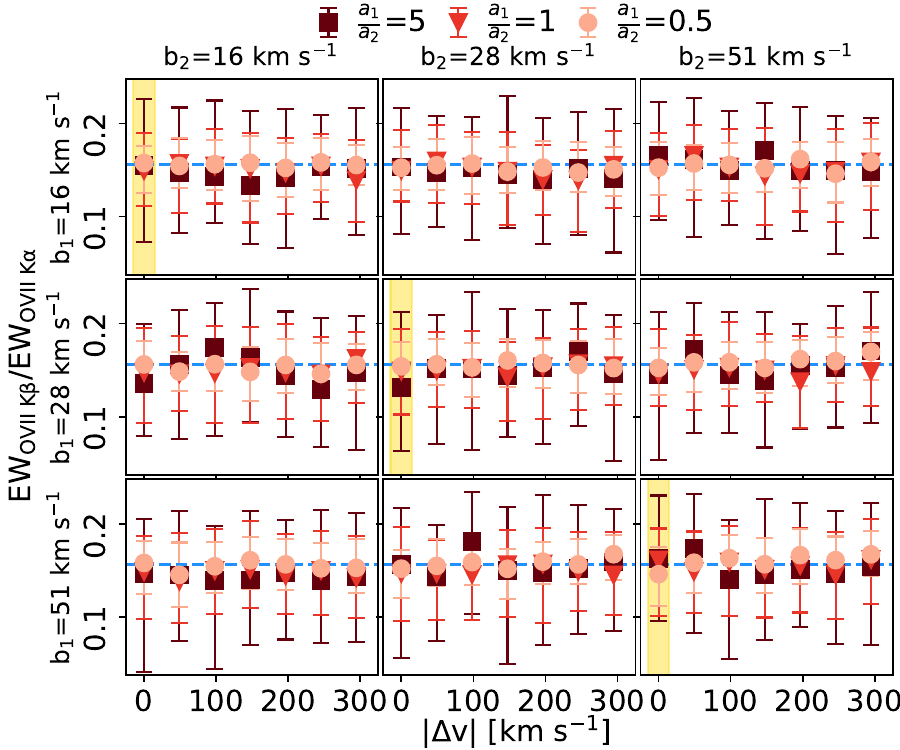}
    \caption{The ratio of best-fitted EW of \oviib and \oviia lines each modeled with 1-Gaussian, as a function of LOS velocity difference, $\rm |\Delta v|$, between the two input Gaussians. Different symbols denote different amplitude ratios of the two input Gaussians (see the legend). The broadening of input Gaussians is mentioned next to each panel. {The results are the same for positive and negative LOS velocity, thus effectively there are 7 $|\rm \Delta v|$ for each combination of line broadening and line amplitude ratio.} The horizontal dashed blue line in each panel corresponds to the EW ratio of unsaturated \ovii lines.}
    \label{fig:disentangle_fit}
\end{figure*}

The line broadening can be \textit{directly} measured in resolved lines. In that case, the effective linewidth might be overestimated if underlying multiple narrow lines at different velocities are modeled with a single broad line. However, in our case of unresolved lines, we cannot directly measure the linewidth. As described in \S\ref{sec:method}, the total broadening is \textit{indirectly} measured from the equivalent widths of two transitions of the same ion. Thus, we need to test whether the true equivalent width is under/overestimated by modeling multiple unresolved lines with a single unresolved line. 

In \texttt{XSPEC}, we simulate (\texttt{fakeit}) mock spectra by modeling the continuum with absorbed power law and \oviia and \oviib transition each as the sum of two unresolved Gaussian absorption lines. Thus the input model is \texttt{tbabs$\times$(powerlaw + zagauss$\rm_{K\alpha,1}$ + zagauss$\rm_{K\beta,1}$ + zagauss$\rm_{K\alpha,2}$ + zagauss$\rm_{K\beta,2}$)}. The flux of the continuum is set sufficiently large to keep \oviib detectable in all spectra. We use the response file and auxiliary file of one of our ACIS/LETG data. 

Each Gaussian (\texttt{zagauss}) is characterized by wavelength ($\lambda$), redshift (z), broadening (b), and amplitude (a). The wavelengths of the Gaussians are set at the $z$=0 wavelengths of \oviia and \oviib transitions. The line-of-sight (LOS) velocity that determines the redshift is kept within the velocity resolution ($\approx$300\,km\,s$^{-1}$) so that it is unresolved from $z$=0. Because \ovii is detectable in the temperature range of $\approx 10^{5.4-6.4}$\,K (see ionization fraction of \ovii in Fig.\,\ref{fig:get_b}, right panel), the line broadening is set to be similar to the thermal broadening in this temperature range, i.e., 16--51 km\,s$^{-1}$. The line broadening and LOS velocity of the two Gaussians of the same transition are allowed to be the same or different. The amplitude of the first Gaussian, $\rm a_1$, is set such that the resulting EW is similar to the median of the measured EW toward 12 sightlines presented in this letter. We consider three amplitudes of the second Gaussian, $\rm a_2$: 1)smaller, 2)equal, and 3)larger than $\rm a_1$. The line broadening and LOS velocities of the two \oviib lines are kept the same as those of the two \oviia lines. The amplitude of \oviib lines is set according to the EW ratio of unsaturated \oviib and \oviia lines (eq.\,1). For each combination of amplitude, line broadening, and LOS velocity of the Gaussian lines, we create 100 mock spectra. The normalized unfolded spectra are shown in Fig.\,\ref{fig:disentangle_sim}. 

  Next, we fit each simulated spectrum in \texttt{XSPEC} by modeling \ovii transitions with one unresolved Gaussian absorption line: \texttt{tbabs$\times$(powerlaw + zagauss$\rm_{K\alpha}$ + zagauss$\rm_{K\beta}$)}. We turn Bayesian inference on and consider the fit statistic of c-stat. The LOS velocity is forced to be within the velocity resolution. The normalization, LOS velocity, and line broadening of \oviib and \oviia lines are allowed to vary independently from each other. We obtain the EW of \ovii lines from the best-fit model for every 100 spectra that were simulated for each combination of input parameters and consider the median and 68\% confidence interval of the EW distribution. The best-fitted EW ratio of the two transitions is shown in Fig.\,\ref{fig:disentangle_fit}. 

The line broadening of the two input Gaussian lines is the same in the diagonal panels of Fig.\,\ref{fig:disentangle_sim} and \ref{fig:disentangle_fit}, i.e., $\rm b_1 = b_2$. Here, the LOS velocity of those two Gaussians is the same in the yellow-shaded region. If two Gaussians of the same width are centered at the same mean value, they are equivalent to one Gaussian with its amplitude equal to the sum of those two Gaussians. Thus, the input models for these cases converge to one Gaussian line, and we can use them for consistency check, i.e., whether we can retrieve the input EW of a single Gaussian and thus verify the performance of the simulation and fit. 

In the non-diagonal panels of Fig.\,\ref{fig:disentangle_sim} and \ref{fig:disentangle_fit}, the input Gaussian with larger line broadening could be treated as a purely thermally broadened line at its input temperature or a non-thermally broadened line at a lower temperature. Thus, these panels serve as lines from multiple temperatures as well as a mix of thermally and non-thermally broadened lines.   

As we see, all EW ratios are consistent with the ratio of unsaturated lines. Thus, fitting two unresolved Gaussians with one unresolved Gaussian is unlikely to result in under/over-estimation of the input EW, irrespective of the LOS velocities and amplitudes of the input Gaussians. Therefore, we rule out that in our data purely thermally broadened multiple components are manifested as a single non-thermally broadened component. Thus, we robustly confirm our detection of non-thermal broadening in the warm-hot CGM.   


\subsection{Multiple temperature components}\label{sec:multiT}
Throughout our analysis, we have assumed that all of the \ovii and \oviii exist in the same phase. Based on the oxygen-only analysis, we cannot test the possibility of deviation from a single phase. However, by simultaneously studying multiple metal ions (e.g., N, Ne, Si) in addition to oxygen, a super-virial $\approx 10^7$\,K hot phase co-existing with the $\approx 10^6$\,K warm-hot CGM has been identified toward several individual high S/N sightlines \citep[highlighted in Fig.\,\ref{fig:vturb}]{Das2019a,Das2021b,McClain2024} and stacking of many low S/N sightlines \citep{Lara2023a}. In those scenarios, \ovii comes predominantly from the warm-hot phase, but \oviii comes from both the hot and warm-hot phases in a comparable amount. Thus, N(\oviiin) in the warm-hot phase is smaller than the total N(\oviiin) measured toward those sightlines. 

In this letter, the temperature is measured from the column density ratio of \oviii and \oviin, and in most of the sightlines, the data is not deep enough to look for other metal lines and test the presence of a super-virial hot phase. A smaller N(\oviiin) would lead to a lower temperature and hence smaller thermal broadening of oxygen lines in the warm-hot phase. As the total line broadening remains unchanged, smaller thermal broadening implies larger non-thermal broadening. For example, the non-thermal line broadening toward Mrk\,421 is $\approx 61$\,km s$^{-1}$ assuming that a single phase contains all of \ovii and \oviiin. However, simultaneous hybrid-ionization modeling of \oviiin-containing warm-hot and hot phases revealed that the non-thermal line broadening of the warm-hot CGM is $\approx 200$\,km s$^{-1}$ \citep{Das2021b}. Even sightlines currently without any evidence of non-thermal broadening in a single-phase scenario might have nonzero non-thermal broadening once they are more accurately characterized in a multiphase scenario with deeper data and rigorous ionization modeling. With the current data, we can qualitatively claim the detection of the effect of non-thermal sources on the warm-hot CGM. 

\section{Summary and conclusions}\label{sec:summary}
In this letter, we present the first detection of the non-thermal line broadening in the warm-hot CGM using $z=0$ X-ray absorption measurements of \ovii and \oviii lines toward quasar sightlines. Theoretical simulations of the warm-hot CGM should include the non-thermal sources to match observations. Below, we summarize our results: 
\begin{enumerate}
    \item Non-thermal line broadening is evident in 8 out of 12 sightlines. On average, the non-thermal broadening is $\approx 50$\% stronger than the thermal broadening. The detection of non-thermal line broadening is not correlated with the column density of \oviin, \oviiin, or the temperature of the warm-hot CGM measured toward these sightlines. 
    \item The correlation between non-thermal line broadening and the longitude of sightlines or the angle from the Galactic center is negligible. Thus, non-thermal sources from nuclear activities, if any, do not have any observable effect on the warm-hot Galactic halo.
    \item Non-thermal line broadening is $\approx 74\%$ stronger toward sightlines at lower galactic latitude. This suggests that non-thermal sources affecting the warm-hot Galactic halo likely originate in the Galactic disk.
\end{enumerate}

We emphasize that we have used archival data of \chandra and \xmmn. All the quasar sightlines were observed to study primarily the target itself or the intervening warm-hot intergalactic medium. Thus, the S/N of the data is not adjusted to the opacity of the warm-hot CGM of the \mwn, leading to inhomogeneous detection sensitivity in the oxygen ions. Despite the limitations, the thermal and chemical characterization of the warm-hot/hot CGM in previous studies and the constraint on the non-thermal broadening in this letter show promising avenues for utilizing high-resolution spectroscopic X-ray data. Deeper observations are required toward sightlines with already detected \oviia line (but no other line) to detect \oviib and \oviii K$\alpha$ lines. It would allow us to improve the estimate of covering fraction and sky distribution of the non-thermal broadening. Surveys targeted towards CGM science with current satellites, e.g., gratings of \chandra and \xmmn, microcalorimeter onboard \textit{XRISM} and proposed missions like \textit{Arcus} and \textit{LEM} would lead to a better understanding of the properties of the warm-hot CGM in X-ray absorption. 
\section{acknowledgments}
{We thank the anonymous referee for constructive comments and suggestions.} S.D. thanks Smita Mathur for the useful discussions. S.D. acknowledges support from the KIPAC Fellowship of Kavli Institute for Particle Astrophysics and Cosmology, Stanford University. This research has made use of NASA's Astrophysics Data System Bibliographic Services. This paper employs a list of \chandra datasets, obtained by the Chandra X-ray Observatory, contained in \dataset[DOI: nonthermal-CGM]{https://doi.org/10.25574/cdc.206}.

\facilities{Chandra, XMM-Newton}

\software{\texttt{CIAO} \citep{Fruscione2006}, \texttt{HeaSoft} \citep{Drake2005}, \texttt{Jupyter} \citep{jupyter2016}, \texttt{Matplotlib} \citep{Hunter2007}, \texttt{NumPy} \citep{numpy2020}, \texttt{SciPy} \citep{scipy2022}, \texttt{XSPEC} \citep{Arnaud1999} }


\begin{thebibliography}{}
\expandafter\ifx\csname natexlab\endcsname\relax\def\natexlab#1{#1}\fi
\providecommand{\url}[1]{\href{#1}{#1}}
\providecommand{\dodoi}[1]{doi:~\href{http://doi.org/#1}{\nolinkurl{#1}}}
\providecommand{\doeprint}[1]{\href{http://ascl.net/#1}{\nolinkurl{http://ascl.net/#1}}}
\providecommand{\doarXiv}[1]{\href{https://arxiv.org/abs/#1}{\nolinkurl{https://arxiv.org/abs/#1}}}

\bibitem[{{Arnaud} {et~al.}(1999){Arnaud}, {Dorman}, \& {Gordon}}]{Arnaud1999}
{Arnaud}, K., {Dorman}, B., \& {Gordon}, C. 1999, {XSPEC: An X-ray spectral
  fitting package}.
\newblock \doeprint{9910.005}

\bibitem[{{Bennett} \& {Sijacki}(2020)}]{Bennett2020}
{Bennett}, J.~S., \& {Sijacki}, D. 2020, \mnras, 499, 597,
  \dodoi{10.1093/mnras/staa2835}

\bibitem[{{Das} {et~al.}(2021){Das}, {Mathur}, {Gupta}, \&
  {Krongold}}]{Das2021b}
{Das}, S., {Mathur}, S., {Gupta}, A., \& {Krongold}, Y. 2021, \apj, 918, 83,
  \dodoi{10.3847/1538-4357/ac0e8e}

\bibitem[{{Das} {et~al.}(2019{\natexlab{a}}){Das}, {Mathur}, {Gupta},
  {Nicastro}, \& {Krongold}}]{Das2019c}
{Das}, S., {Mathur}, S., {Gupta}, A., {Nicastro}, F., \& {Krongold}, Y.
  2019{\natexlab{a}}, \apj, 887, 257, \dodoi{10.3847/1538-4357/ab5846}

\bibitem[{{Das} {et~al.}(2019{\natexlab{b}}){Das}, {Mathur}, {Nicastro}, \&
  {Krongold}}]{Das2019a}
{Das}, S., {Mathur}, S., {Nicastro}, F., \& {Krongold}, Y. 2019{\natexlab{b}},
  \apjl, 882, L23, \dodoi{10.3847/2041-8213/ab3b09}

\bibitem[{{Draine}(2011)}]{Draine2011}
{Draine}, B.~T. 2011, {Physics of the Interstellar and Intergalactic Medium}
  (Princeton University Press)

\bibitem[{{Drake}(2005)}]{Drake2005}
{Drake}, S.~A. 2005, in X-Ray and Radio Connections, ed. L.~O. {Sjouwerman} \&
  K.~K. {Dyer} (Santa Fe, New Mexico: NRAO), 6.01

\bibitem[{{Fang} {et~al.}(2015){Fang}, {Buote}, {Bullock}, \& {Ma}}]{Fang2015}
{Fang}, T., {Buote}, D., {Bullock}, J., \& {Ma}, R. 2015, \apjs, 217, 21,
  \dodoi{10.1088/0067-0049/217/2/21}

\bibitem[{{Fielding} {et~al.}(2020){Fielding}, {Tonnesen}, {DeFelippis}, {Li},
  {Su}, {Bryan}, {Kim}, {Forbes}, {Somerville}, {Battaglia}, {Schneider}, {Li},
  {Choi}, {Hayward}, \& {Hernquist}}]{Fielding2020}
{Fielding}, D.~B., {Tonnesen}, S., {DeFelippis}, D., {et~al.} 2020, \apj, 903,
  32, \dodoi{10.3847/1538-4357/abbc6d}

\bibitem[{{Fruscione} {et~al.}(2006){Fruscione}, {McDowell}, {Allen},
  {Brickhouse}, {Burke}, {Davis}, {Durham}, {Elvis}, {Galle}, {Harris},
  {Huenemoerder}, {Houck}, {Ishibashi}, {Karovska}, {Nicastro}, {Noble},
  {Nowak}, {Primini}, {Siemiginowska}, {Smith}, \& {Wise}}]{Fruscione2006}
{Fruscione}, A., {McDowell}, J.~C., {Allen}, G.~E., {et~al.} 2006, in Society
  of Photo-Optical Instrumentation Engineers (SPIE) Conference Series, Vol.
  6270, Society of Photo-Optical Instrumentation Engineers (SPIE) Conference
  Series, 62701V, \dodoi{10.1117/12.671760}

\bibitem[{{Gommers} {et~al.}(2022){Gommers}, {Virtanen}, {Burovski},
  {Weckesser}, {Oliphant}, {Cournapeau}, {Haberland}, {Reddy}, {Alexbrc},
  {Peterson}, {Nelson}, {Wilson}, {Endolith}, {Mayorov}, {Polat}, {Van Der
  Walt}, {Laxalde}, {Brett}, {Larson}, {Millman}, {Lars}, {Peterbell10}, {Roy},
  {Van Mulbregt}, {Carey}, {Eric-Jones}, {Sakai}, {Moore}, {Kai}, \&
  {Kern}}]{scipy2022}
{Gommers}, R., {Virtanen}, P., {Burovski}, E., {et~al.} 2022, {scipy/scipy:
  SciPy 1.8.0}, v1.8.0, Zenodo,  Zenodo, \dodoi{10.5281/zenodo.595738}

\bibitem[{{Gupta} {et~al.}(2023){Gupta}, {Mathur}, {Kingsbury}, {Das}, \&
  {Krongold}}]{Gupta2023}
{Gupta}, A., {Mathur}, S., {Kingsbury}, J., {Das}, S., \& {Krongold}, Y. 2023,
  Nature Astronomy, 7, 799, \dodoi{10.1038/s41550-023-01963-5}

\bibitem[{{Gupta} {et~al.}(2012){Gupta}, {Mathur}, {Krongold}, {Nicastro}, \&
  {Galeazzi}}]{Gupta2012}
{Gupta}, A., {Mathur}, S., {Krongold}, Y., {Nicastro}, F., \& {Galeazzi}, M.
  2012, \apjl, 756, L8, \dodoi{10.1088/2041-8205/756/1/L8}

\bibitem[{Harris {et~al.}(2020)Harris, Millman, van~der Walt, Gommers,
  Virtanen, Cournapeau, Wieser, Taylor, Berg, Smith, Kern, Picus, Hoyer, van
  Kerkwijk, Brett, Haldane, del R{\'{i}}o, Wiebe, Peterson,
  G{\'{e}}rard-Marchant, Sheppard, Reddy, Weckesser, Abbasi, Gohlke, \&
  Oliphant}]{numpy2020}
Harris, C.~R., Millman, K.~J., van~der Walt, S.~J., {et~al.} 2020, Nature, 585,
  357, \dodoi{10.1038/s41586-020-2649-2}

\bibitem[{{Heesen} {et~al.}(2019){Heesen}, {Whitler}, {Schmidt}, {Miskolczi},
  {Sridhar}, {Horellou}, {Beck}, {G{\"u}rkan}, {Scannapieco}, {Br{\"u}ggen},
  {Heald}, {Krause}, {Paladino}, {Nikiel-Wroczy{\'n}ski}, {Wilber}, \&
  {Dettmar}}]{Heesen2019}
{Heesen}, V., {Whitler}, L., {Schmidt}, P., {et~al.} 2019, \aap, 628, L3,
  \dodoi{10.1051/0004-6361/201936046}

\bibitem[{{Heesen} {et~al.}(2023){Heesen}, {O'Sullivan}, {Br{\"u}ggen}, {Basu},
  {Beck}, {Seta}, {Carretti}, {Krause}, {Haverkorn}, {Hutschenreuter},
  {Bracco}, {Stein}, {Bomans}, {Dettmar}, {Chy{\.z}y}, {Heald}, {Paladino}, \&
  {Horellou}}]{Heesen2023}
{Heesen}, V., {O'Sullivan}, S.~P., {Br{\"u}ggen}, M., {et~al.} 2023, \aap, 670,
  L23, \dodoi{10.1051/0004-6361/202346008}

\bibitem[{{Henley} {et~al.}(2010){Henley}, {Shelton}, {Kwak}, {Joung}, \& {Mac
  Low}}]{Henley2010}
{Henley}, D.~B., {Shelton}, R.~L., {Kwak}, K., {Joung}, M.~R., \& {Mac Low},
  M.-M. 2010, \apj, 723, 935, \dodoi{10.1088/0004-637X/723/1/935}

\bibitem[{{HI4PI Collaboration} {et~al.}(2016){HI4PI Collaboration}, {Ben
  Bekhti}, {Fl{\"o}er}, {Keller}, {Kerp}, {Lenz}, {Winkel}, {Bailin},
  {Calabretta}, {Dedes}, {Ford}, {Gibson}, {Haud}, {Janowiecki}, {Kalberla},
  {Lockman}, {McClure-Griffiths}, {Murphy}, {Nakanishi}, {Pisano}, \&
  {Staveley-Smith}}]{Bekhti2016}
{HI4PI Collaboration}, {Ben Bekhti}, N., {Fl{\"o}er}, L., {et~al.} 2016, \aap,
  594, A116, \dodoi{10.1051/0004-6361/201629178}

\bibitem[{{Hunter}(2007)}]{Hunter2007}
{Hunter}, J.~D. 2007, Computing in Science and Engineering, 9, 90,
  \dodoi{10.1109/MCSE.2007.55}

\bibitem[{{Ji} {et~al.}(2020){Ji}, {Chan}, {Hummels}, {Hopkins}, {Stern},
  {Kere{\v{s}}}, {Quataert}, {Faucher-Gigu{\`e}re}, \& {Murray}}]{Ji2020}
{Ji}, S., {Chan}, T.~K., {Hummels}, C.~B., {et~al.} 2020, \mnras, 496, 4221,
  \dodoi{10.1093/mnras/staa1849}

\bibitem[{Kluyver {et~al.}(2016)Kluyver, Ragan-Kelley, P{\'e}rez, Granger,
  Bussonnier, Frederic, Kelley, Hamrick, Grout, Corlay, Ivanov, Avila, Abdalla,
  \& Willing}]{jupyter2016}
Kluyver, T., Ragan-Kelley, B., P{\'e}rez, F., {et~al.} 2016, in Positioning and
  Power in Academic Publishing: Players, Agents and Agendas, ed. F.~Loizides \&
  B.~Schmidt, IOS Press, 87 -- 90.
\newblock \url{https://ebooks.iospress.nl/publication/42900}

\bibitem[{{Lara-DI} {et~al.}(2023){Lara-DI}, {Mathur}, {Krongold}, {Das}, \&
  {Gupta}}]{Lara2023a}
{Lara-DI}, A.~J., {Mathur}, S., {Krongold}, Y., {Das}, S., \& {Gupta}, A. 2023,
  \apj, 946, 55, \dodoi{10.3847/1538-4357/acbf40}

\bibitem[{{Mathur}(2022)}]{Mathur2022}
{Mathur}, S. 2022, in Handbook of X-ray and Gamma-ray Astrophysics. Edited by
  Cosimo Bambi and Andrea Santangelo, 59,
  \dodoi{10.1007/978-981-16-4544-0_112-1}

\bibitem[{{McClain} {et~al.}(2024){McClain}, {Mathur}, {Das}, {Krongold}, \&
  {Gupta}}]{McClain2024}
{McClain}, R.~L., {Mathur}, S., {Das}, S., {Krongold}, Y., \& {Gupta}, A. 2024,
  \mnras, 527, 5093, \dodoi{10.1093/mnras/stad3497}

\bibitem[{{Nicastro} {et~al.}(2016){Nicastro}, {Senatore}, {Krongold},
  {Mathur}, \& {Elvis}}]{Nicastro2016b}
{Nicastro}, F., {Senatore}, F., {Krongold}, Y., {Mathur}, S., \& {Elvis}, M.
  2016, \apj, 828, L12, \dodoi{10.3847/2041-8205/828/1/L12}

\bibitem[{{Nicastro} {et~al.}(2002){Nicastro}, {Zezas}, {Drake}, {Elvis},
  {Fiore}, {Fruscione}, {Marengo}, {Mathur}, \& {Bianchi}}]{Nicastro2002}
{Nicastro}, F., {Zezas}, A., {Drake}, J., {et~al.} 2002, \apj, 573, 157,
  \dodoi{10.1086/340489}

\bibitem[{{Schmidt} {et~al.}(2021){Schmidt}, {Schmidt}, \&
  {Grete}}]{Schmidt2021}
{Schmidt}, W., {Schmidt}, J.~P., \& {Grete}, P. 2021, \aap, 654, A115,
  \dodoi{10.1051/0004-6361/202140920}

\bibitem[{{Spitzer}(1956)}]{Spitzer1956}
{Spitzer}, Lyman, J. 1956, \apj, 124, 20, \dodoi{10.1086/146200}

\bibitem[{{Truong} {et~al.}(2023){Truong}, {Pillepich}, {Nelson}, {Bogd{\'a}n},
  {Schellenberger}, {Chakraborty}, {Forman}, {Kraft}, {Markevitch},
  {Ogorzalek}, {Oppenheimer}, {Sarkar}, {Veilleux}, {Vogelsberger}, {Wang},
  {Werner}, {Zhuravleva}, \& {Zuhone}}]{Truong2023}
{Truong}, N., {Pillepich}, A., {Nelson}, D., {et~al.} 2023, \mnras, 525, 1976,
  \dodoi{10.1093/mnras/stad2216}

\bibitem[{{van de Voort} {et~al.}(2021){van de Voort}, {Bieri}, {Pakmor},
  {G{\'o}mez}, {Grand}, \& {Marinacci}}]{Voort2021}
{van de Voort}, F., {Bieri}, R., {Pakmor}, R., {et~al.} 2021, \mnras, 501,
  4888, \dodoi{10.1093/mnras/staa3938}

\bibitem[{{Williams} {et~al.}(2005){Williams}, {Mathur}, {Nicastro}, {Elvis},
  {Drake}, {Fang}, {Fiore}, {Krongold}, {Wang}, \& {Yao}}]{Williams2005}
{Williams}, R.~J., {Mathur}, S., {Nicastro}, F., {et~al.} 2005, \apj, 631, 856,
  \dodoi{10.1086/431343}

\end{thebibliography}
\bibliographystyle{aasjournal}

\end{document}